\documentclass[aps,prl,superscriptaddress,nofootinbib,twocolumn]{revtex4-1}
\usepackage{mathptmx}

\usepackage[T1]{fontenc}
\usepackage[latin9]{inputenc}
\usepackage{amsmath}
\usepackage{graphicx}
\makeatletter

\newcommand{\lyxmathsym}[1]{\ifmmode\begingroup\def\b@ld{bold}
  \text{\ifx\math@version\b@ld\bfseries\fi#1}\endgroup\else#1\fi}

\makeatother

\usepackage{babel}
\begin{document}

\title{Scattering Signatures of Bond-Dependent Magnetic Interactions}
\author{Joseph A. M. Paddison}
\email{paddisonja@ornl.gov}

\affiliation{Materials Science and Technology Division, Oak Ridge National Laboratory,
Oak Ridge, TN 37831, USA}
\begin{abstract}
Bond-dependent magnetic interactions can generate exotic phases
such as Kitaev spin-liquid states. Experimentally determining
the values of bond-dependent interactions is
a challenging but crucial problem. Here, I show that each symmetry-allowed
nearest-neighbor interaction on triangular and honeycomb lattices
has a distinct signature in paramagnetic neutron-diffraction data,
and that such data contain sufficient information to determine
the spin Hamiltonian unambiguously \emph{via} unconstrained fits.
Moreover, I show that bond-dependent interactions can often be extracted
from powder-averaged data. These results facilitate experimental determination
of spin Hamiltonians for materials that do not show conventional magnetic ordering.
\end{abstract}
\maketitle
The discovery and characterization of magnetic materials with novel ground states such as topological order is an overarching goal of condensed-matter physics.
Such materials have potential applications for topological quantum
computation \citep{Kitaev_2003,Nayak_2008}, and are of fundamental
interest because they can show entangled ground states whose excitations
have fractional quantum numbers \citep{Takagi_2019,Broholm_2020}.
Traditionally, the search for such states has concentrated on materials
with isotropic (Heisenberg) magnetic interactions. However, the discovery
of the celebrated Kitaev model \citep{Kitaev_2003,Baskaran_2007,Hermanns_2018,Rousochatzakis_2018}---in which bond-dependent interactions on the honeycomb lattice stabilize a spin-liquid
ground state with fractionalized excitations---has led to intense interest
in materials where strong spin-orbit coupling generates bond-dependent interactions \citep{Jackeli_2009,Chaloupka_2010,Winter_2017,Sano_2018}.
Candidate honeycomb-lattice materials include $\alpha$-RuCl$_{3}$ \citep{Plumb_2014,Banerjee_2016,Do_2017,Banerjee_2017,Winter_2017a},
YbCl$_{3}$ \citep{Sala_2019}, NaNi$_{2}$BiO$_{6-\delta}$ \citep{Scheie_2019},
H$_{3}$LiIr$_{2}$O$_{6}$ \citep{Kitagawa_2018,Yadav_2018}, Na$_{2}$IrO$_{3}$
\citep{Singh_2012,Rau_2014,Hwan-Chun_2015}, and $\alpha$-Li$_{2}$IrO$_{3}$
\citep{Williams_2016,Choi_2019}. Bond-dependent interactions on the
triangular lattice may generate quantum spin-liquid
states \citep{Zhu_2018}, with potential realizations including YbMgGaO$_{4}$
\citep{Li_2015,Shen_2016,Paddison_2017,Li_2017a}, NaYbS$_{2}$ \citep{Baenitz_2018,Sarkar_2019},
and NaYbO$_{2}$ \citep{Bordelon_2019,Ding_2019}.

Robust experimental determination of bond-dependent interactions is
key to identifying promising candidate materials. Yet,
such interactions are challenging to measure; e.g., in the well-studied
Kitaev candidate material $\alpha$-RuCl$_{3}$, no clear consensus
has been reached on the sign or magnitude of the Kitaev interaction \citep{Laurell_2020}.
There are two main reasons for such difficulties. First, the spin Hamiltonian
for triangular and honeycomb lattices contains four nearest-neighbor
interactions \citep{Chaloupka_2015}, but most experiments are sensitive
only to a subset of these. Second, current data-analysis approaches
typically assume conventional long-range magnetic ordering---e.g., to
model magnon spectra \citep{Banerjee_2016,Ran_2017,Ozel_2019,Zhang_2018,Ross_2011}---but
such ordering is not
expected in topologically-ordered or spin-liquid states \citep{Broholm_2020}. When long-range
ordering does occur in candidate materials, it is often unclear if
it is driven by the nearest-neighbor model or by perturbations
such as further-neighbor interactions or structural disorder
\citep{Lampen-Kelley_2017,Zhu_2017,Li_2017,Sarkar_2020,Winter_2016}.

In this Letter, I explore the extent to which bond-dependent interactions
can be extracted from neutron-diffraction patterns measured in the
\emph{paramagnetic} phase, above any spin ordering or freezing temperature
$T_{N}$. Such data show a continuous (diffuse) variation of the magnetic
scattering intensity $I(\mathbf{Q})$ with wavevector
$\mathbf{Q}=h\mathbf{a}^{*}+k\mathbf{b}^{*}+l\mathbf{c}^{*}$. Crucially, the diffuse $I(\mathbf{Q})$ varies continuously with the underlying magnetic interactions and so may, in principle, determine them uniquely; however, previous modeling focused on bond-independent interactions \cite{Blech_1964,Manuel_2009,Fennell_2009,Bai_2019, Samarakoon_2020}.
By contrast, Bragg diffraction below $T_N$ only restricts the interactions to a (frequently large) search space compatible with the observed ordering \cite{Rau_2014}. 
I proceed by simulating diffuse $I(\mathbf{Q})$ data for classical bond-dependent
models (test cases) on triangular and honeycomb lattices. I show that
such data contain signatures of the signs of bond-dependent
interactions, the interaction values can be accurately determined
\emph{via} unconstrained fits to simulated data, and this approach is robust to statistical noise typical of real measurements. Perhaps most surprisingly, the powder averaged
$I(Q=|\mathbf{Q}|)$ retains some sensitivity to bond-dependent
interactions, and so can constrain
them when single-crystal samples are unavailable. 

\begin{figure}
\centering{}\includegraphics{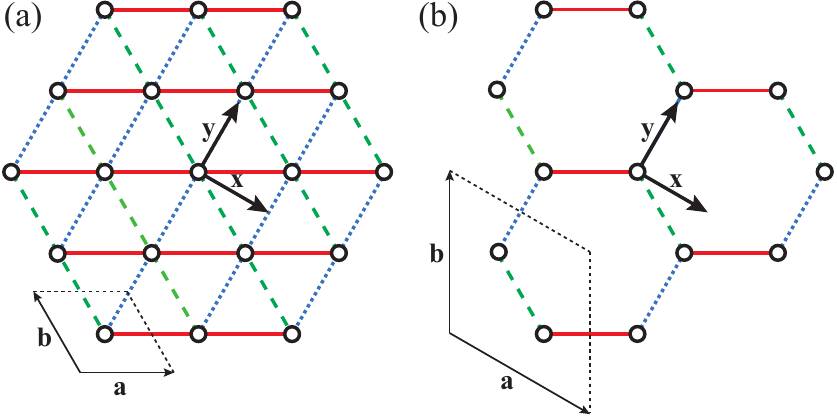}\caption{\label{fig:fig1} (a) Triangular and (b) honeycomb lattices. Spins
are referred to Cartesian axes $\mathbf{x}$, $\mathbf{y}$, and $\mathbf{z}$, with $\mathbf{z}$ directed out of the page.
Conventional unit-cell vectors are $\mathbf{a}$, $\mathbf{b}$,
and $\mathbf{c}\parallel\mathbf{z}$. The three bond types are
shown as solid red, dashed green, and dotted blue lines.}
\end{figure}

The most general nearest-neighbor spin Hamiltonian allowed by threefold
symmetry of the magnetic site has the same form for triangular and
honeycomb lattices \cite{Rau_2018},
\begin{align}\label{eq:hamiltonian}
\mathcal{H}& =\sum_{\left\langle i,j\right\rangle }\Bigl\{ J_{X}\left(S_{i}^{x}S_{j}^{x}+S_{i}^{y}S_{j}^{y}\right)+J_{Z}S_{i}^{z}S_{j}^{z}\nonumber \\
 & +J_{A}\left[(S_{i}^{x}S_{j}^{x}-S_{i}^{y}S_{j}^{y})\cos\phi_{ij}-(S_{i}^{x}S_{j}^{y}+S_{i}^{y}S_{j}^{x})\sin\phi_{ij}\right]\nonumber \\
 & -J_{B}\sqrt{2}\left[(S_{i}^{x}S_{j}^{z}+S_{i}^{z}S_{j}^{x})\cos\phi_{ij}+(S_{i}^{y}S_{j}^{z}+S_{i}^{z}S_{j}^{y})\sin\phi_{ij}\right]\Bigr\},
\end{align}
where superscript $x$, $y$, and $z$ denote spin components with
respect to the $\mathbf{x}$, $\mathbf{y}$, and $\mathbf{z}$ axes
shown in Fig.~\ref{fig:fig1}, and $\phi_{ij}\in\left\{ \frac{2\pi}{3},-\frac{2\pi}{3},0\right\} $
for bonds colored red, green, and blue respectively in Fig.~\ref{fig:fig1}.
The Hamiltonian contains four interactions, whose physical origin
is typically superexchange between trigonally-distorted edge-sharing
$M$O$_{6}$ octahedra \citep{Rau_2014}: $J_{X}$ and $J_{Z}$ describe
a conventional $XXZ$ model, while $J_{A}$ and $J_{B}$ are bond
dependent. Several parameterizations of Eq.\,(\ref{fig:fig1}) are
in use \citep{supp}; I follow the conventions of Ref.~\citealp{Chaloupka_2015},
which resemble those applied to YbMgGaO$_{4}$ \citep{Li_2015,Shen_2016,Paddison_2017,Li_2017a}.
A different parameterization $\{J,K,\Gamma,\Gamma^{\prime}\}$ is
typically used for honeycomb systems \citep{Rau_2014,Chaloupka_2015}.
However, we will see that Eq.~(\ref{eq:hamiltonian}) has advantages
for interpreting $I(\mathbf{Q})$ data.

\begin{figure*}
\centering{}\includegraphics{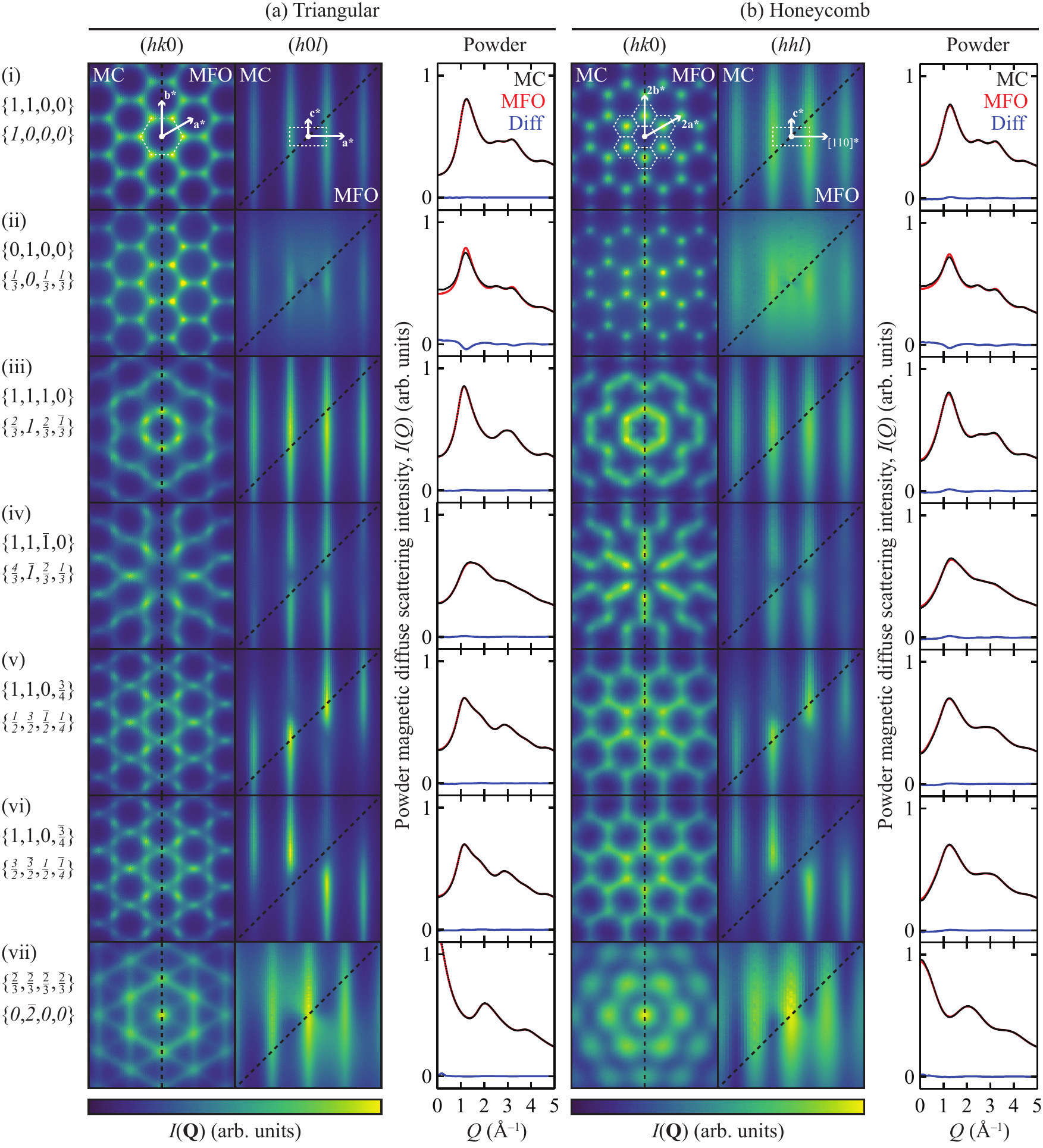}\caption{\label{fig:fig2} Simulated neutron-diffraction patterns in the paramagnetic
phase for test cases (i)--(vii) discussed in the text. The interaction
parameters for each model are shown left, with $\{J_{X},J_{Z},J_{A},J_{B}\}$
in roman type above the corresponding $\{J,K,\Gamma,\Gamma^{\prime}\}$
in \emph{italics}. Column (a) shows calculations for the triangular
lattice (left to right: $(hk0)$, $(h0l)$, and powder diffraction
patterns) and column (b) shows calculations for the honeycomb lattice
(left to right: $(hk0)$, $(hhl)$, and powder diffraction patterns). An isotropic $g$-factor is assumed. 
Results for Monte Carlo (MC) and reaction-field (MFO) approximations
are shown separated by dashed black lines, as labeled on the top panels.
For powder patterns, MC results are shown as black circles; MFO results
as red lines; and difference (MC--MFO) as blue lines. All powder
patterns are shown on the same intensity scale. For single-crystal
patterns, reciprocal-lattice vectors $\mathbf{a}^{*}$, $\mathbf{b}^{*}$,
and $\mathbf{c}^{*}$ are labeled in the top panels, and the first
Brillouin zone is shown as a white dashed line. Both single-crystal
planes are shown on the same intensity scale for each test case except
honeycomb (i) and (ii), for which the intensity scale is doubled in
the $(hhl)$ plane for clarity. In all calculations, the triangular
unit cell has dimensions $|\mathbf{a}|=|\mathbf{b}|=3.464\,\lyxmathsym{\protect\AA}$,
$|\mathbf{c}|=6.0\,\lyxmathsym{\protect\AA}$, and the honeycomb unit
cell has dimensions $|\mathbf{a}|=|\mathbf{b}|=|\mathbf{c}|=6.0\,\lyxmathsym{\protect\AA}$.}
\end{figure*}

I consider seven test cases with interaction parameters $\{J_{X},J_{Z},J_{A},J_{B}\}$
(``$J$'s'') covering a range of interaction space: (i) the antiferromagnetic
(AF) Heisenberg model, $\{1,1,0,0\}$; (ii) the AF Ising model, $\{0,1,0,0\}$;
(iii, iv) the AF Heisenberg model with $J_{A}=+1$ and $-1$ ($\equiv\bar{1})$,
respectively; (v, vi) the AF Heisenberg model with $J_{B}=+\frac{3}{4}$
and $-\frac{3}{4}$, respectively; and (vii) the ferromagnetic Kitaev
model, $\{\bar{\frac{2}{3}},\bar{\frac{2}{3}},\bar{\frac{2}{3}},\bar{\frac{2}{3}}\}$,
which corresponds to $\{J,K,\Gamma,\Gamma^{\prime}\}=\{0,\bar{2},0,0\}$.
Test cases (i) and (ii) are not bond dependent and are included for
comparison; (iii)--(vi) explore the effect of changing signs of bond-dependent
terms and are potentially relevant to YbMgGaO$_{4}$ \citep{Li_2015,Shen_2016,Paddison_2017,Li_2017a};
and (vii) explores the Kitaev limit potentially relevant to $\alpha$-RuCl$_{3}$
\citep{Banerjee_2016,Do_2017,Banerjee_2017,Winter_2017a}. A further 20 test cases, corresponding to models proposed for $\alpha$-RuCl$_{3}$ \cite{Laurell_2020}, are considered in SM \cite{supp}.
For test cases (i)--(vii), I performed classical Monte Carlo (MC) simulations of Eq.~(\ref{eq:hamiltonian})
with spin length $|\mathbf{S}|=1$ \citep{supp}. The simulation temperature
$T=2$ (in the same units as the $J$'s) for (iii)--(vii) on the
triangular lattice, and $T=1$ otherwise, which is well above $T_{N}$
in all cases. The energy-integrated magnetic neutron-diffraction intensity
\begin{align}
I(\mathbf{Q}) & \propto[f(Q)]^{2}\sum_{i,j,\alpha,\beta}p_{\alpha\beta}\bigl\langle S_{i}^{\alpha}S_{j}^{\beta}\bigr\rangle e^{i\mathbf{Q}\cdot\mathbf{r}_{ij}},\label{eq:ns}
\end{align}
where $\alpha,\beta\in\left\{ x,y,z\right\} $ denote Cartesian components,
$\mathbf{r}_{ij}$ is the vector connecting spins $i$ and $j$, $f(Q)$
denotes an arbitrary magnetic form factor (Yb$^{3+}$) \citep{Brown_2004},
and
\begin{equation}
p_{\alpha\beta}\equiv\delta_{\alpha\beta}-Q_{\alpha}Q_{\beta}/Q^{2}\label{eq:projection}
\end{equation}
is the projection factor \citep{Lovesey_1987,Halpern_1939,Paddison_2019},
which arises because neutrons only ``see'' spin components perpendicular
to $\mathbf{Q}$, and couples spin and spatial degrees of freedom.
Eq.~(\ref{eq:projection}) is key to magnetic crystallography because
it usually allows the absolute spin structure to be solved 
from $T<T_N$ neutron-diffraction data \citep{Shirane_1959}. I will show that it
also allows bond-dependent interactions to be inferred from $T>T_N$ neutron-diffraction data.

Fig.~\ref{fig:fig2} shows the single-crystal $I(\mathbf{Q})$ and
powder $I(Q)$ \cite{Blech_1964} for all test cases. Two orthogonal single-crystal planes
are shown: $(hk0)$, and either $(h0l)$ for the triangular lattice
or $(hhl)$ for the honeycomb lattice. Our first key result is that
$I(\text{\textbf{Q})}$ is qualitatively different in each case. In
particular, it is strongly affected by changing the sign of $J_{A}$
or $J_{B}$, whereas other experiments (e.g.,
magnon spectra \citep{Banerjee_2016,Ran_2017,Ozel_2019,Zhang_2018})
are usually insensitive to at least one of these signs. The differences in the plane perpendicular to $(hk0)$
do not arise from inter-layer interactions---absent in all test cases---but
instead from the projection factor, as I now discuss for each test
case. (i) The Heisenberg diffraction pattern repeats periodically, except for the trivial decrease of intensity with
$f(Q)$. This is because all diagonal correlators $\langle S_{i}^{\alpha}S_{j}^{\alpha}\rangle$
are equal and all off-diagonal correlators $\langle S_{i}^{\alpha}S_{j}^{\beta}\rangle$
are zero; hence $\langle p_{\alpha\alpha}\rangle=2/3$ is independent
of $\mathbf{Q}$. (ii) The Ising diffraction pattern repeats periodically in the $(hk0)$ plane but shows further $\mathbf{Q}$-dependence
in the perpendicular plane, because the intensity is dominated
by $p_{zz}\langle S_{i}^{z}S_{j}^{z}\rangle=(1-Q_{z}^{2}/Q^{2})\langle S_{i}^{z}S_{j}^{z}\rangle$
terms. (iii, iv) Nonzero $J_{A}$ causes nontrivial $\mathbf{Q}$-dependence
in both planes because it drives nonzero $\langle S_{i}^{x}S_{j}^{y}\rangle$
and $\langle S_{i}^{y}S_{j}^{x}\rangle$ correlators, so that terms
like $p_{xy}\langle S_{i}^{x}S_{j}^{y}\rangle=-Q_{x}Q_{y}\langle S_{i}^{x}S_{j}^{y}\rangle/Q^{2}$
contribute to $I(\mathbf{Q})$. (v, vi) Nonzero $J_{B}$ also causes
nontrivial $\mathbf{Q}$-dependence in both planes, but unlike the
previous cases, $I(hkl)\neq I(hk\bar{l})$. This is because nonzero
$J_{B}$ lowers the hexagonal symmetry of the previous models to trigonal
\citep{Chaloupka_2015}, yielding nonzero terms like $p_{xz}\langle S_{i}^{x}S_{j}^{z}\rangle$
and $p_{yz}\langle S_{i}^{y}S_{j}^{z}\rangle$ that change sign under
either $(hkl)\rightarrow(hk\bar{l})$ or\textbf{ $S_{i}^{z}\rightarrow-S_{i}^{z}$}
for all $S^{z}$. Since the latter is equivalent to $J_{B}\rightarrow-J_{B}$
in Eq.~(\ref{eq:hamiltonian}), both $(hkl)\rightarrow(hk\bar{l})$
and $J_{B}\rightarrow-J_{B}$ have the same effect on $I(\mathbf{Q})$.
These results follow from basic properties of Eqs.~(\ref{eq:hamiltonian})--(\ref{eq:projection})
that apply for quantum as well as classical systems, and show that
each interaction has a different effect on $I(\mathbf{Q})$. Dominant interactions can therefore be identified by inspection
of diffuse-scattering data.

I now obtain a theory that explains the modulation of $I(\mathbf{Q})$.
I employ the Onsager reaction-field (MFO) method \citep{Brout_1967,Wysin_2000}
previously shown to give accurate results for Heisenberg models \citep{Logan_1995,Eastwood_1995,Hohlwein_2003,Conlon_2010,Bai_2019,Plumb_2019}.
The Fourier transform of the interactions $J_{ij}^{\alpha\beta}(\mathbf{Q})\equiv-\sum_{\mathbf{R}}J_{ij}^{\alpha\beta}(\mathbf{R})e^{-\mathrm{i}\mathbf{Q}\cdot\mathbf{R}}$,
where $J_{ij}^{\alpha\beta}(\mathbf{R})$ is the coefficient of $S_{i}^{\alpha}S_{j}^{\beta}$
in Eq.\,(\ref{eq:hamiltonian}) for sites $i$ and $j$ separated
by a lattice vector $\mathbf{R}$. The $J_{ij}^{\alpha\beta}(\mathbf{Q})$
are elements of a $3N\times3N$ interaction matrix, where $N$ is
the number of sites in the unit cell. For the triangular lattice ($N=1$), the interaction matrix
\begin{equation}
\mathsf{J}(\mathbf{Q})=-\left(\begin{array}{ccc}
aJ_{X}+bJ_{A} & cJ_{A} & -\sqrt{2}bJ_{B}\\
cJ_{A} & aJ_{X}-bJ_{A} & \sqrt{2}cJ_{B}\\
-\sqrt{2}bJ_{B} & \sqrt{2}cJ_{B} & aJ_{Z}
\end{array}\right),\label{eq:interaction_matrix}
\end{equation}
where $a=2[\cos 2\pi(h+k)+\cos 2\pi h+\cos 2\pi k]$, $b=2\cos 2\pi (h+k)-\cos 2\pi h-\cos 2\pi k$,
and $c=\sqrt{3}(\cos 2\pi k-\cos 2\pi h)$. For the honeycomb lattice ($N=2$), the interaction
matrix
\begin{equation}
\mathsf{J}_{\mathrm{h}}(\mathbf{Q})=\left(\begin{array}{cc}
0 & \mathsf{J}\\
\mathsf{J}^{*} & 0
\end{array}\right),
\end{equation}
where $a$, $b$, and $c$ in Eq.~(\ref{eq:interaction_matrix})
are replaced by $a_{\mathrm{h}}=1+e^{2\pi\mathrm{i}h}+e^{-2\pi\mathrm{i}k}$,
$b_{\mathrm{h}}=e^{-2\pi\mathrm{i}k}-(1+e^{2\pi\mathrm{i}h})/2$, and
$c_{\mathrm{h}}=\sqrt{3}(1-e^{2\pi\mathrm{i}h})/2$, respectively.
Diagonalizing the interaction matrix at each $\mathbf{Q}$ yields
its eigenvalues $\lambda_{\mu}$ and eigenvector components $U_{\mu}^{\alpha,i}$,
where $\mu$ labels the $3N$ eigenmodes and $i$ labels sites at
positions $\mathbf{r}_{i}$ in the unit cell. The $T>T_{N}$ scattering
intensity in the reaction-field approximation is given by
\begin{equation}
I_{\mathrm{MFO}}(\mathbf{Q})\propto\frac{[f(Q)]^{2}}{3N}\sum_{\mu=1}^{3N}\frac{|\mathbf{s}_{\mu}(\mathbf{Q})|^{2}}{1-\chi_{0}(\lambda_{\mu}(\mathbf{Q})-\lambda)},\label{eq:intensity_onsager}
\end{equation}
where $\chi_{0}=1/3T$ is the Curie susceptibility, and $\mathbf{s}_{\mu}(\mathbf{Q})=\sum_{i,\alpha}(\hat{\mathbf{n}}_{\alpha}-\mathbf{Q}\thinspace\hat{\mathbf{n}}_{\alpha}\cdot\mathbf{Q}/Q^{2})U_{\mu}^{\alpha,i}e^{\mathrm{i}\mathbf{Q}\cdot\mathbf{r}_{i}}$
with $\hat{\mathbf{n}}_{\alpha}\in\{\mathbf{x},\mathbf{y},\mathbf{z}\}$.
Eq.~(\ref{eq:intensity_onsager}) is identical to the mean-field
expression \citep{Enjalran_2004} except for the reaction field $\lambda$,
which is determined self-consistently by requiring that $\sum_{\mu,\mathbf{q}}[1-\chi_{0}(\lambda_{\mu}(\mathbf{q)}-\lambda)]^{-1}=3NN_{\mathbf{q}}$
for a grid of $N_{\mathbf{q}}=40^{3}$ wavevectors in the Brillouin
zone. Fig.~\ref{fig:fig2} compares the single-crystal $I(\mathbf{Q})$
and powder $I(Q)$ from reaction-field theory with the accurate MC
results. The agreement is very good in all cases; only in the Ising
case are subtle differences evident. The success of reaction-field theory for bond-dependent interactions is
remarkable given its simplicity.

\begin{figure}
\centering{}\includegraphics{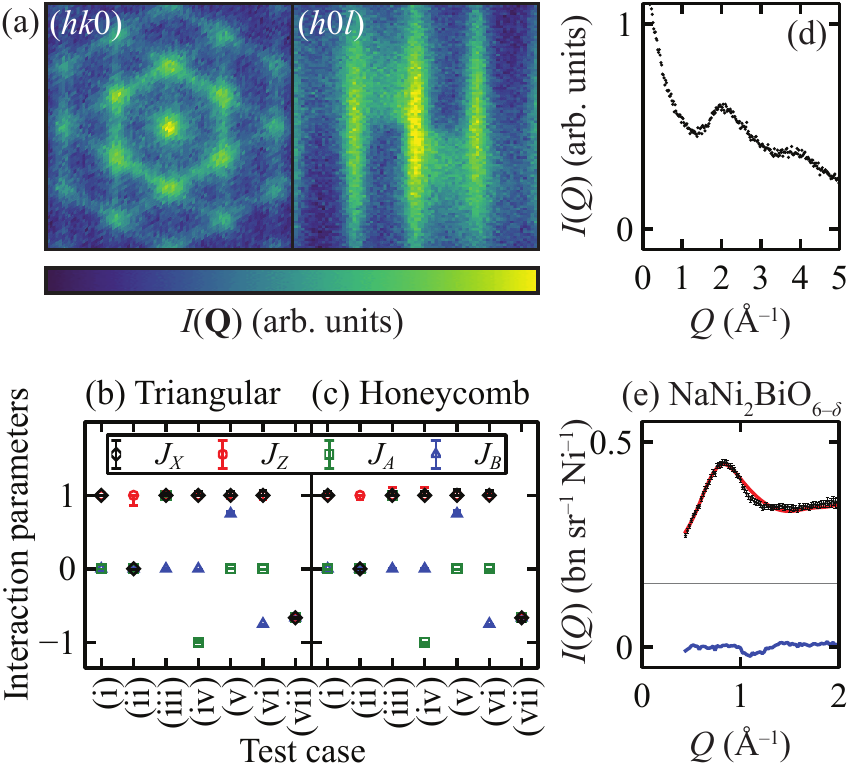}\caption{\label{fig:fig3}
(a) Simulated ``noisy'' single-crystal data $I(\mathbf{Q})$ with
5\% error bars for the Kitaev model
on the triangular lattice (test case (vii)).
(b,c) Values of the interaction parameters for test cases
(i)--(vii) for (b) triangular and (c) honeycomb lattices. In each
case, $J_{X}$ is shown as black diamonds, $J_{Z}$ as red circles,
$J_{A}$ as green squares, and $J_{B}$ as blue triangles. Error bars
indicate values obtained from unconstrained fits of all four parameters
to two single-crystal $I(\mathbf{Q})$ planes, such as those shown in (a).
(d) Simulated ``noisy'' powder $I(Q)$ data with
1\% error bars for test case (vii). (e) Experimental $I(Q)$ data for NaNi$_2$BiO$_{6-\delta}$ (black circles), fit (red line), data--fit (blue line), and fitted incoherent level (grey line). }
\end{figure}

The sensitivity of $I(\mathbf{Q})$ to bond-dependent interactions
suggests that it may be possible to solve the inverse problem---to
infer interaction values from $I(\mathbf{Q})$ data. To test this
possibility, I performed unconstrained fits of the four $J$'s, using MC single-crystal scattering planes as simulated ``data'' for
each test case. To make the tests more realistic, data were adulterated with with random noise drawn from a normal distribution with $\sigma$ equal to 5\% of the maximum intensity
(``5\% error bars''), as shown in Fig.~\ref{fig:fig3}(a).
An intensity scale factor was also fitted, as required if data are not normalized in absolute intensity
units. In the fits, $I(\mathbf{Q})$ was calculated
in the reaction-field approximation because it is computationally
efficient and free from statistical noise. The nonlinear least-squares
algorithm in the \textsc{Minuit} program \citep{James_1975} was used
to minimize the sum of squared residuals $\chi^{2}$. If the $J$'s
are fully determined by the data, a fit should converge to a global
minimum $\chi_{\mathrm{min}}^{2}$ with nearly correct $J$'s, provided the initial $J$'s are sufficiently close to optimal.
Conversely, if the $J$'s are underdetermined, fits will either fail, or yield several different solutions with indistinguishable fit quality depending on initial $J$'s.
A unique solution is defined here as the absence of low-lying false minima with $\chi^{2}<\chi_{\mathrm{min}}^{2}+15$, where this condition reflects the 99\% confidence interval for five parameters \citep{James_1994}.
To test for uniqueness, I performed 50 separate
fits initialized with different $J$'s randomly distributed in the
range $\{-0.5:0.5\}$ \cite{Senn_2016}. In every test case, the fits identified a unique solution with nearly correct $J$'s, and convergence was achieved
from nearly all (96\%) of the initial parameter sets. Similarly favorable results were obtained for 20 $\alpha$-RuCl$_3$ test cases \cite{supp}, demonstrating that the approach is robust to inclusion of a third-neighbor interaction and the rapid decay of the Ru$^{3+}$ magnetic form factor \cite{Do_2017}.
Fig.~\ref{fig:fig3}(b,c)
shows the systematic error in the optimal $J$'s due to the inaccuracy
of the reaction-field approximation. This error is usually small
and the worst-case error is $0.14$ in $J_{Z}$. These results show
that bond-dependent interactions can be reliably extracted from noisy and unnormalized
$I(\mathbf{Q})$ data.

As a more challenging test, I considered \emph{powder-averaged} $I(Q)$ data with 1\% error bars {[}Fig.~\ref{fig:fig3}(d){]}. On
the one hand, powder averaging causes much information loss. In particular,
powder data cannot distinguish $\pm J_{B}$, because $J_{B}\rightarrow-J_{B}$
is equivalent to $(hkl)\rightarrow(hk\bar{l})$; I therefore consider
test cases (v, vi) together. On the other hand, $I(Q)$ differs for
the other test cases {[}Fig.~\ref{fig:fig2}{]}. Remarkably, fits of the
four $J$'s to noisy $I(Q)$ data yielded a unique optimal solution
with nearly correct $J$'s in 10 out of 12 test cases.
In the remaining cases---(iii) and (v, vi) for the triangular
lattice---two different solutions were identified, which had nearly the
same $\chi^{2}$. Parameter uncertainties were also increased compared
to single-crystal fits \citep{supp}. Despite these limitations, the
ability of powder fits to identify a small number of candidate models
suggests that $I(Q)$ can provide a ``fingerprint'' of bond-dependent
interactions---a compact data set that contains most of the discriminating
information.

I finally apply this methodology to published neutron data of the candidate Kitaev material NaNi$_2$BiO$_{6-\delta}$ ($\delta=0.33$) \cite{Scheie_2019}, in which Ni$^{3+}$ ions ($S=3/2$, $J=1/2$) occupy a honeycomb lattice. 
The experimental $I(Q)$ data shown in Fig.~\ref{fig:fig3}(e) were obtained by energy-integrating the $T=10$\,K ($>T_\mathrm{N}$) inelastic neutron-scattering data of Ref.~\onlinecite{Scheie_2019}. In the fits, the measured magnetic moment of $2.21(1)$\,$\mu_\mathrm{B}$ per Ni$^{3+}$ was assumed \cite{Scheie_2019}, and an incoherent (flat-in-$Q$) signal was fitted. For all fits, the magnitude of $K$ is at least twice that of $J$, $\Gamma$, and $\Gamma^{\prime}$, and the predicted in-plane magnetic ordering wavevector $\mathbf{k}\approx({\frac{1}{3},\frac{1}{3}})$ is consistent with the measured value \cite{Scheie_2019}. These results demonstrate the successful application of our methodology to experimental data and support the dominant Kitaev interactions proposed in NaNi$_2$BiO$_{6-\delta}$ \cite{Scheie_2019}.

These results show that bond-dependent interactions on triangular
and honeycomb lattices have signatures in diffuse neutron-scattering
data at $T>T_{N}$ that enable estimation of the interactions\emph{
via} unconstrained fits. This unexpected sensitivity is mainly due
to the projection factor, Eq.~(\ref{eq:projection}); hence, it is
important to measure $I(\mathbf{Q})$ outside the $(hk0)$ plane where
this factor is significant, and to include it in calculations, which has not often been done. Our
methodology is generally applicable and employs conventional least-squares
optimization \citep{Bai_2019}, providing a robust and computationally-efficient
alternative to machine-learning-based approaches \citep{Samarakoon_2020},
as well as to interaction-independent approaches such as reverse Monte
Carlo refinement \citep{Paddison_2012} and pair-distribution-function
analysis \citep{Frandsen_2014}. Key advantages are that measurements in high magnetic fields are not required, and additional data such as bulk magnetic susceptibility---related to $I(Q\rightarrow0)$ \cite{Marshall_1968}---can be included. 
A limitation is that quantum effects
that redistribute scattering intensity \citep{Mourigal_2013,Samarakoon_2017}
are not included: this may cause inaccuracy in fitted interaction
values, but does not affect sensitivity to interaction signs. Moreover, a fit typically
requires only a few hundred $I(\mathbf{Q})$ calculations for convergence---taking
$\sim$$60$\,s to fit to $\sim$$10^{4}$ data points on a laptop---so
that replacement of classical calculations by more-expensive quantum
calculations is feasible. If interlayer spin correlations are negligible above $T_{N}$, our results are unaffected by the layer
stacking sequence---a useful feature because of the prevalence of stacking
faults in quasi-2D materials \citep{Johnson_2015}. 
These results promise to accelerate experimental determination of spin Hamiltonians
of candidate materials that do not exhibit conventional magnetic ordering, such as in the emerging field
of ``topology by design'' metal-organic frameworks \citep{Yamada_2017}.

I am grateful to Xiaojian Bai (ORNL), Andrew Christianson (ORNL), Seung-Hwan Do (ORNL), Mechthild Enderle (ILL), Andrew Goodwin (Oxford), Pontus Laurell (ORNL), Martin Mourigal (Georgia Tech), Kate Ross (Colorado State), Allen Scheie (ORNL), Ross Stewart (ISIS), and Alan Tennant (ORNL) for valuable discussions. I also thank the authors of Ref.~\onlinecite{Scheie_2019} for making available their published data. This work was supported by the Laboratory Directed Research and Development Program of Oak Ridge National Laboratory, managed by UT-Battelle, LLC for the US Department of Energy (project design, calculations, and manuscript writing), and by the U.S. Department of Energy, Office of Science, Basic Energy Sciences, Materials Sciences and Engineering Division (computational resources). I acknowledge a Junior Research Fellowship from Churchill College, University of Cambridge, during which computer programs underlying this work were written.

\end{document}